\documentclass[iop]{emulateapj}

\usepackage{apjfonts}
\usepackage{epsfig}
\usepackage{epstopdf}
\usepackage{natbib}
\usepackage{amssymb,latexsym}
\usepackage{graphicx}
\usepackage{amsmath}

\usepackage[usenames]{color}

\shortauthors{Reddick}
\shorttitle{Russell Conjecture}

\begin{document}
\title{Empirical Limits on the Russell Conjecture}
\author{Rachel Reddick}

\begin{abstract}
The Russell Conjecture states that there is an unproven possibility of small (<1 m) hollow heat-resistant objects (HoHOs) in Earth orbit or otherwise present in the inner solar system or asteroid belt.  While such objects are not the current target of any ongoing searches, we can place stringent limits on their presence using current optical and infrared surveys.  The high albedo of HoHOs partially compensates for their small size.  As such, we find that no HoHOs greater than 10 cm in radius to a distance of at least 30,000 km, by the Air Force Space Surveillance System.  Objects of that size in a stable orbit at 384,000 km (the Earth-Moon distance) may be detected and confirmed by more infrequent, deeper sweeps of the same system.  However, it remains possible for undetected HoHOs to exist in near-Earth or Martian orbit.  We discuss the prospects of finding such HoHOs in the near future with new telescopes such as LSST.
\end{abstract}


\maketitle

\section{Introduction}

The Russell Conjecture, first postulated by \cite{Russell1952}, states that "If I were to suggest that between the Earth and Mars there is a china teapot revolving about the sun in an elliptical orbit, nobody would be able to disprove my assertion provided I were careful to add that the teapot is too small to be revealed even by our most powerful telescopes."  Colloquially known as Russell's teapot or the celestial teapot, a hollow heat-resistant object of this nature may now be observable due to significant improvements in telescope capabilities.

The Air Force Space Surveillance System is a space radar system currently capable of detecting objects 10 cm in size to a distance of at least 30,000 km \cite{w-AFSSS}.  Further upgrades are proposed, though these focus primarily on improving the detection rate of space junk in low Earth orbit \citep{spacefence}.

However, the smallest near-Earth objects detected which are not in local orbit are still approximately a few meters in size (\citet{nearmiss,J2009}).  As such, there are few precise limits on the presence of hollow heat-resistant objects (HoHOs) in the solar system.  The remained of this paper will examine the expected properties of HoHOs, 

\section{Expected Properties}

Numerous artificial HoHOs are produced on Earth, though precise rates are not reported.   Estimating  100 million teapots exist, with a typical lifetime of ten years, suggests an approximate worldwide production of 10 million teapots.  It is conceivable that a small fraction of these might make their way into orbit, either due to ejection from an impactor or due to some kind of strange accident at the International Space Station.

\begin{figure*}
\includegraphics[width=0.3\textwidth]{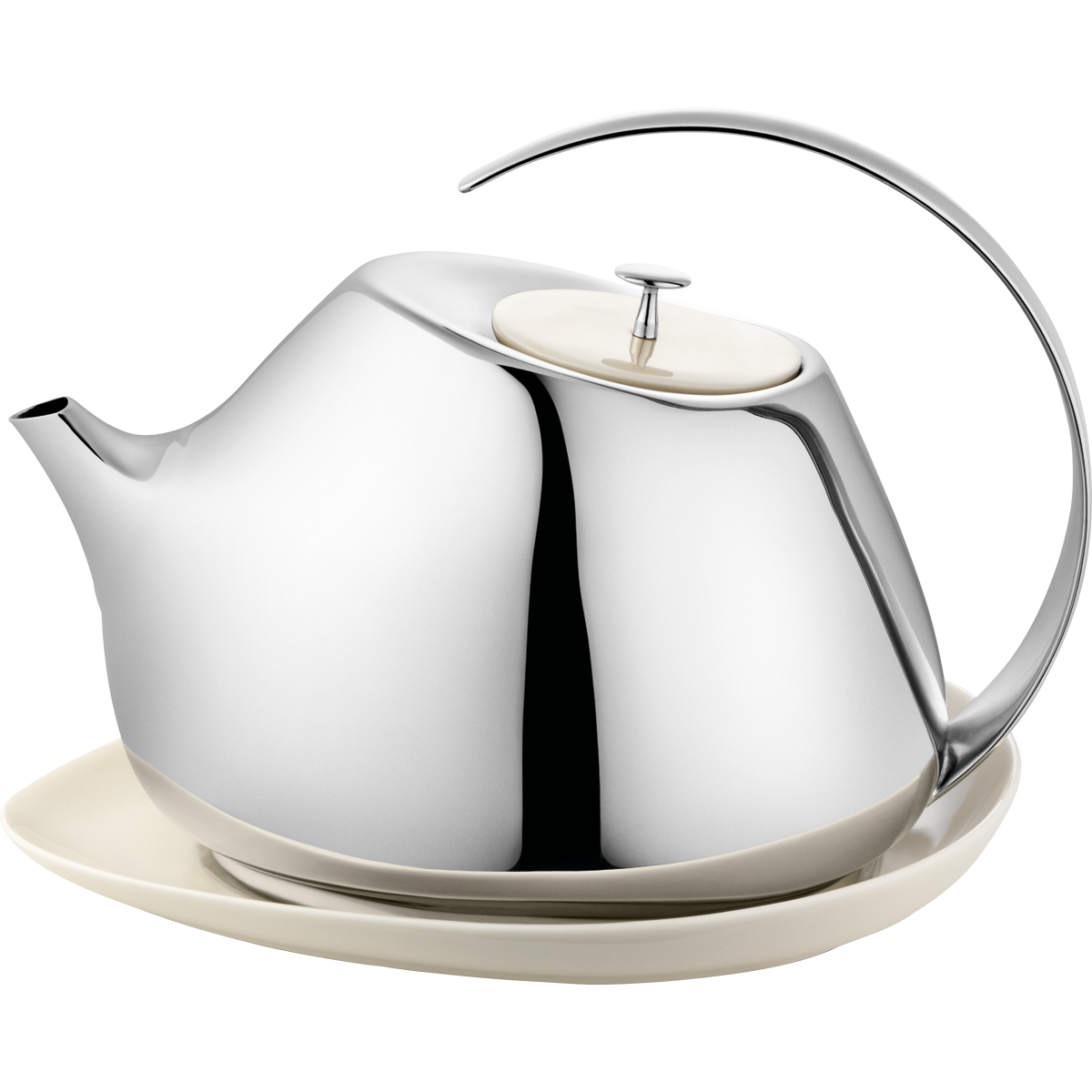}
\includegraphics[width=0.3\textwidth]{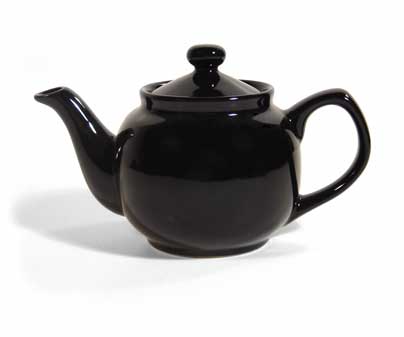}
\includegraphics[width=0.3\textwidth]{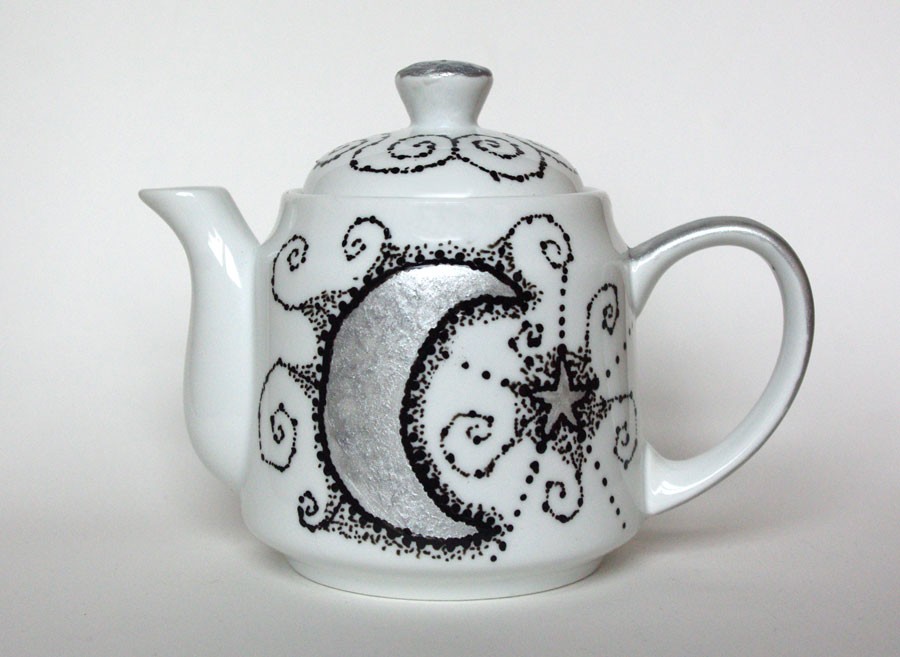}
\caption{High resolution images of several possible types of HoHOs.  Chocolate teapots are excluded as being insufficiently durable to have a significant lifetime in space \citep{w-teapot}.}
\label{fig:teapots}
\end{figure*}

As most terrestrial HoHOs are composed of high reflectivity ceramics, glasses or metals, we anticipate that most extraterrestrial HoHOs to have similar properties.  Given that most radar operates with wavelengths of a few centimeters, they will be possible but difficult to pick up using typical planetary radar techniques.  Assuming a typical cross section of  approximately $80~\rm{cm}^2$, a HoHO with an albedo of 1 will reflect 10 W of sunlight when orbiting at a distance of 1 AU from the Sun.  As viewed from Earth at a distance of 1 AU, the HoHO will have an apparent magnitude of approximately 37 (absolute magnitude H=37), which is so dim it's not even funny.  Given that the upcoming Large Synoptic Survey Telescope will only observe to a depth of about magnitude 25 each night \citep{lsst}, a HoHO will need to be closer than 0.005 AU, or less than three times the Earth-Moon distance.

Given their high albedos, HoHOs are expected to cool rapidly.  This is especially applicable to the high-albedo class of objects.  As such, we expect that detection of objects with infrared telescopes such as the Wide-Field Infrared Survey Explorer (WISE) to also be very challenging \citep{wise}.

\section{Current Observations and Possible Impactors}

\begin{figure}
\includegraphics[width=0.5\textwidth]{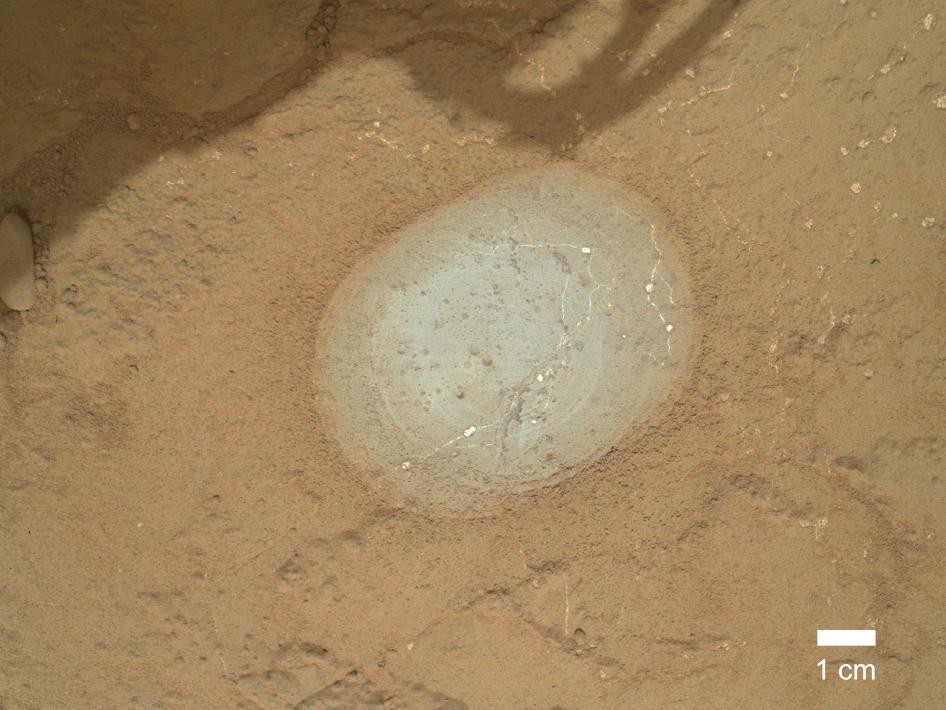}
\caption{A section of rock uncovered by the NASA Curiosity rover following brushing away of surface dust.  It kind of looks like it might be ceramic, although it isn't really.  Image credit to NASA.}
\label{fig:rover}
\end{figure}

Due to their hollow and fragile nature, we expect that most HoHOs will burn up on atmospheric entry on Earth.  However, it is possible that HoHOs or pieces of them can survive entering the thinner Martian atmosphere and survive to hit the surface.  Rumors of a possible kaolinite fragment found on Mars the the NASA Curiosity rover, originating from a ceramic-based HoHO, are completely ridiculous and unfounded, but we're still hopeful.  (See Fig.~\ref{fig:rover})

Over the lifetime of the solar system, we expect that the population of HoHOs would be gradually depleted by impact onto planets and also by interactions with other asteroids or HoHOs.  As a result, there may be relatively few HoHOs intact at the present time.  Nonetheless, we expect that fragments may be found associated with larger objects or the zodiacal dust.

\section{Related Object Types}

Terrestrial HoHOs are commonly found in association with water and cured leaves of the plant {\em Camella sinensis}.

If such a tea-based asteroid exists, we expect that the most similar terrestrial tea is likely of the pu-erh variety.  This type of tea has been fermented over a long period of time, and is general served in solid balls or pieces broken off from a larger slab.  Pu-erh teas are only typically fermented on Earth for two to ten years prior to consumption.  A tea-based asteroid has probably been fermenting for a time on the order of billions of years, comparable to the solar system age.  The process would be slowed by cold and vacuum.  Overall, we expect that the ultimate result would be almost, but not quite, entirely unlike tea.  Composition is likely similar to chondritic asteroids.  Reports that fragments of the Chelyabinsk meteorite turned a small Russian lake into a weak black tea have yet to be confirmed.

An asteroid large enough to contain all the tea in China would have a mass of 1.5 million tons.  This would have a volume of roughly $1.5\cdot10^6~\rm{m}^3$, far larger than the Tunguska meteor or the Chelyabinsk meteor \citep{chelya}.  With a radius of approximately 140 m, this is large enough to be found by LSST or other next-generation surveys.

\section{Discussion and Conclusions}

We conclude that it remains a very difficult task to place precise limits on the Russell Conjecture, although near-Earth orbits are strongly excluded.

We leave the origins of solar system HoHOs to future work.

\section*{Acknowledgements}

The author would like to thank Michael Busch, as well as Michael Shaw and other Stanford graduate students for helpful discussions.

This article has totally not been endorsed by Stanford University, SLAC, or any other scientific institution.

The author is not responsible for any stupid claims you make make as a consequence of reading this article, including those due to failing to check the date.

\bibliography{russell.bib}

\end{document}